\documentstyle[aps,prl,draft,twocolumn,floats]{revtex}

\def\Teff{T_{\text{eff}}}
\def\be{\begin{equation}}
\def\ee{\end{equation}}
\def\Re{\text{Re}}

\input epsf.sty

\begin{document}

\draft

\wideabs{

\title{Observation of Fluctuation-Dissipation-Theorem Violations 
in a Structural Glass}
\author{Tom\'as~S.\ Grigera$^*$  and  N.~E.\ Israeloff$^\dag$}
\address{Department of Physics and Center for Interdisciplinary
Research on Complex Systems, Northeastern University,\\
 Boston, MA 02115}
\date{November 13, 1999}
\maketitle

\begin{abstract}

The fluctuation-dissipation theorem (FDT), connecting dielectric
susceptibility and polarization noise was studied in glycerol below
its glass transition temperature $T_g$.  Weak FDT violations were
observed after a quench from just above to just below $T_g$, for
frequencies above the $\alpha$ peak.  Violations persisted up to
$10^5$ times the thermal equilibration time of the configurational
degrees of freedom under study, but comparable to the average
relaxation time of the material. These results suggest that excess
energy flows from slower to faster relaxing modes.

\end{abstract}

\pacs{PACS 61.43.Fs 64.70.Pf }

}

When a system is in thermodynamic equilibrium, its linear response
functions (e.g.\ susceptibility) and its equilibrium fluctuations are
related by the well-known fluctuation-dissipation theorem
(FDT)\cite{kubo}. Special cases of the FDT include Nyquist's
formula\cite{nyquist} relating electrical resistance to voltage noise,
and Einstein's relation\cite{einstein}, which links the diffusion and
friction coefficients in Brownian motion.

There is no reason to expect the FDT to hold for systems out of
equilibrium, and it is indeed known to fail in far-from-equilibrium
cases such as Barkhausen noise\cite{barkhausen}. Glasses, however, are
systems that are out of equilibrium but evolve very slowly. Much has
been learned by studying how response functions {\em age,} that is,
depend on the {\em waiting time} ($t_w$) elapsed since a temperature
quench\cite{age}.

Recent theoretical work has shown that the FDT is violated by several
mean-field spin glass models below some critical
temperature\cite{leticia93}, and also that this violation can be used
to define an effective temperature\cite{leticia97}. Computer
simulations have also found FDT violations in finite-range spin
glasses\cite{franz95marinari98}, domain growth
processes\cite{barrat98}, some models of structural
glasses\cite{parisi97barratkob98}, elastic strings in random
media\cite{yoshino98}, kinetically constrained lattice
gases\cite{sellito98} and Ising models with dipolar
interaction\cite{stariolo99}. In all cases, violations are found when
the characteristic observation time, $t$, is of the same order or
greater than the age of the system, {\sl i.e.\/} when $t \gtrsim t_w$
or $\omega t_w \sim 1$, where $\omega \sim 1/t$ is a measurement
frequency.

There is considerable interest in studying FDT violations
experimentally, in part because predicted violations are
model-dependent\cite{leticia98}.  Measurements of FDT violations would
thus furnish a new way to test models of the glass transition and
glassy dynamics. Moreover, recent theoretical results\cite{franz98}
surprisingly indicate that non-equilibrium FDT violations can, at
least in spin glasses, give information about an {\em equilibrium}
order parameter, which would be otherwise inaccessible.

Aging in susceptibility\cite{leheny98} and noise\cite{israeloff98} has
been measured in structural glasses.  However, measurements of the
{\em relation} between noise and response thus far reported in
structural\cite{nathan96} as well as spin\cite{ocio85} glasses are
compatible with the FDT within experimental errors.  This gap between
theory and experiment may be due to the fact that the regime $t_w
\omega \sim 1$ is very difficult to access and/or that actual
violations are rather weak.

In this letter we present results that show a weak violation of the
FDT in a supercooled liquid, glycerol, following a quench below its
glass transition temperature. The effective temperature ($\Teff$)
measured via dielectric noise slowly relaxes towards the thermal bath
temperature $T$. Most surprisingly, violations are observed even with
$t_w \omega \gg 1$. The relaxation is so slow compared to the thermal
equilibration time of the configurational degrees of freedom (CDF)
probed by the experiment, that these results can only be understood in
terms of strong coupling between CDF.

\begin{figure}
\epsfxsize=3.375 in \epsfbox{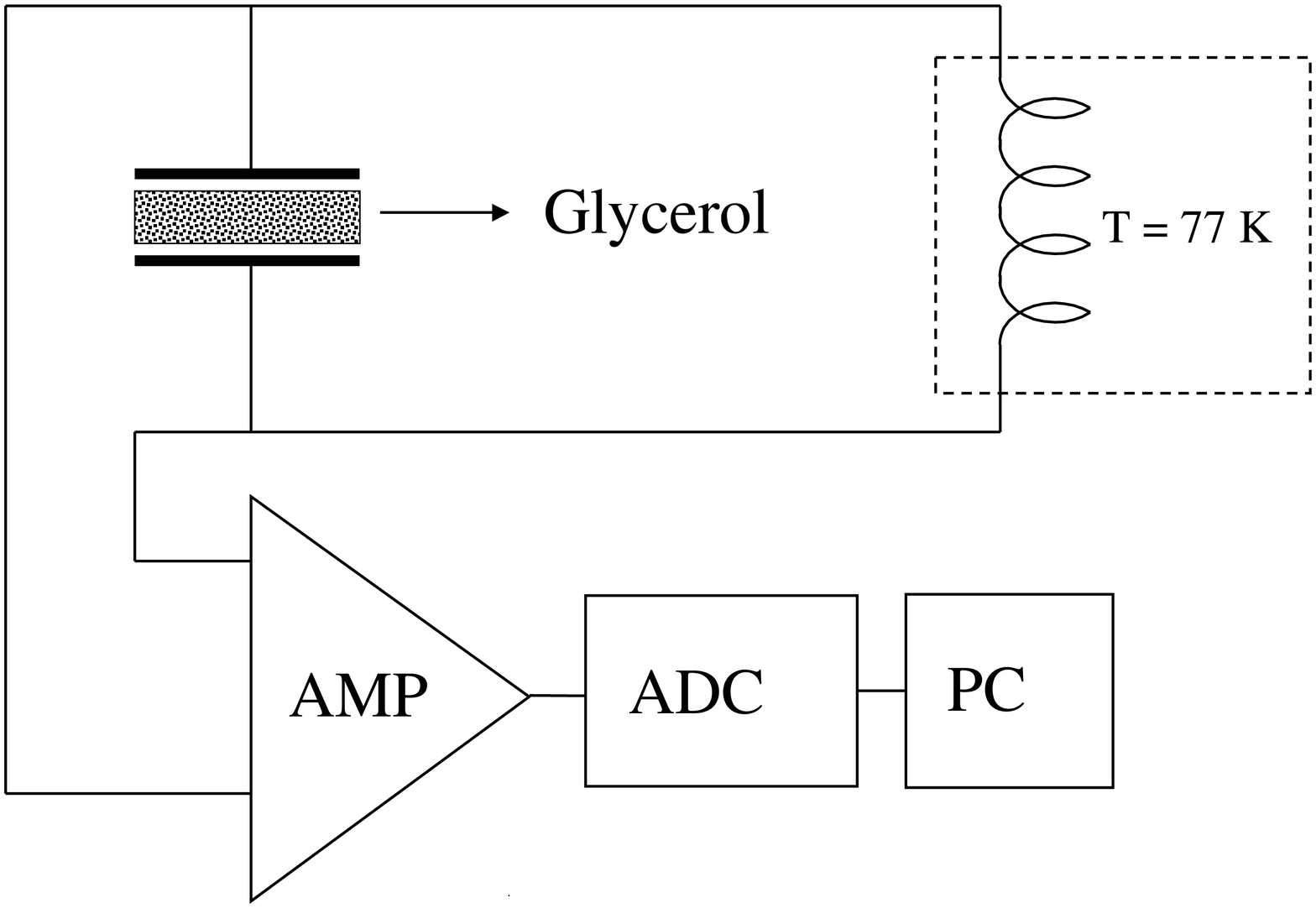}
\caption{Experimental setup. The sample forms the dielectric of a capacitor
in a resonant circuit, which is driven by the polarization
noise of the sample.}
\label{setup}
\end{figure}

  The experiment is similar to the ``oscillator as thermometer''
gedanken experiment proposed by Cugliandolo {\sl et al.\
}\cite{leticia97}.  The idea is to connect a capacitor, whose
dielectric is the glass under study, in parallel with a lossless
inductor (see fig.~\ref{setup}). In thermodynamic equilibrium, the
average energy of the oscillator, $C\langle V^2 \rangle$, will be $k_B
T$ by equipartition, where $\langle V^2 \rangle$ is the integrated
voltage noise, $k_B$ Boltzmann's constant and $T$ the temperature.
The noise power spectral density, $S_V$, will be given by Nyquist's
formula\cite{nyquist},
\be
  S_V(\omega) = \frac{2}{\pi} k_B T \Re Z \label{SV1} 
              = \frac{2}{\pi} k_B T \frac{ - \omega^3 L^2 C''} 
	          { (1- \omega^2 L C')^2 + \omega^4 L^2 C''^2}.
\ee
Here $Z$ is the circuit impedance, $L$ the inductance, and
$C=C'+iC''=C_0 \epsilon$ the capacitance (with $\epsilon$ the complex
dielectric susceptibility).

\begin{figure}
\epsfxsize=3.375 in \epsfbox{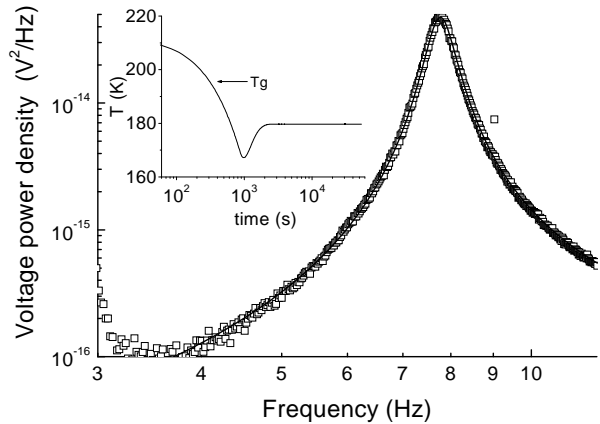}
\caption{Typical voltage power spectrum for $179.8\,$K (squares),
and fit to eq.~\protect \ref{PSfit} (solid line). 
Measurement are made after bath temperature $T$ has stabilized.
Inset: Typical thermal history.}
\label{PS} \label{Thist}
\end{figure}

The strongly peaked noise spectrum produced by the resonant circuit
(see fig.~\ref{PS}) provides {\em three} independent quantities: peak
position, width and height, which relate directly to $C'$, $C''$ and
$T$, respectively. If the FDT ceases to be valid, then this would be
apparent as a failure of the noise spectrum to satisfy Nyquist's
equation with $T$ as the thermodynamic bath temperature. Thus, the
need for separate measurement of noise and response is eliminated with
this method. We note that measurement of broadband noise
alone\cite{israeloff98} only provides two quantities (spectral density
and exponent) and therefore is insufficient to test FDT violations.

For a non-equilibrium (aging) glass, the FDT is believed to
fail\cite{leticia93}, and an effective temperature,
$\Teff(\omega,t_w)$, can be defined\cite{leticia97}. It should replace
$T$ in eq.~\ref{SV1}, control the direction of heat flow, and
determine the average energy of an oscillator coupled to the sample
(as in the equipartition theorem). $\Teff(\omega,t_w)$ should approach
the bath temperature for $t_w\to\infty$ at any fixed frequency, but
its behavior is otherwise system-dependent.  Whether this scenario is
completely valid or not, we can use $\Teff$ to quantify any observed
FDT violations.

The sample capacitor was fabricated by rolling two copper foils
separated by a sheet of high porosity paper on a copper tube, to which
a heater and temperature sensor were attached. Liquid glycerol was
then absorbed into the paper. The capacitor was connected in parallel
with an inductor and both placed inside a liquid nitrogen cryostat at
low pressure ($\approx 100\,$mTorr).  To minimize the noise due to the
lossy part of the inductance, the inductor was placed at the bottom of
the cryostat in good thermal contact with the liquid nitrogen. The
voltage noise was amplified with a Stanford SR560 preamplifier,
filtered with a Stanford SR640 low-pass filter, digitally sampled with
a personal computer and Fourier-analyzed to obtain the voltage power
spectrum as a function of aging time.

Since capacitor and inductor are at different temperatures,
eq.~\ref{SV1} has to be replaced by a generalized
formula\cite{bennet}, which for our circuit gives
\be
 S_V(\omega) = \frac{2k_B}{\pi} \frac{ -\omega^3 \Teff(\omega_0,t_w) |L|^2 C'' 
                                    - \omega T_0 L''}
 {1 - 2\omega^2 (L'C' - L''C'') + \omega^4 |L|^2 |C|^2},
 \label{PSfit}
\ee
where $T_0=77\,$K is the inductor temperature, which now has a lossy
part ($L=L'+iL''$), which is measured separately with a known
capacitor. We disregard any possible $\omega$ dependence of $\Teff$
because $S_V$ is strongly peaked, and its value at the resonance
frequency, $\omega_0$, will dominate.  All other quantities being
known, a least-squares fit of the power spectra to eq.~\ref{PSfit}
yields $C'$, $C''$ and $\Teff$.

The sample was always heated to a bath temperature of $210\,$K (above
the glycerol glass temperature) and then cooled to $T$ near
$180\,$K. A typical thermal history is shown in fig.~\ref{Thist}
(there is an unavoidable temperature undershoot). All reported results
are from data taken after the bath temperature stabilized at its final
value. $t_w=0$ was taken as the time when the sample falls out of
equilibrium. This depends on cooling rate and thermal history. We have
determined $T_g$ to be $\approx 196\,$K.  Adjustment of $T_g$ by a few
degrees will shift the time axis by at most a few hundred seconds,
without affecting our results.

Fig.~\ref{PS} shows a typical power spectrum, together with the fit
using eq.~\ref{PSfit}. The spectrum is an average obtained from
voltage noise sampled in a time window around a particular $t_w$. The
time window increased with $t_w$, so as to reduce errors for large
$t_w$. For shorter $t_w$ ($<10^5\,$s), it was necessary to average
data from several ($\sim 15$) runs.  The resonance frequency is $f_0
\approx 7.7\,$Hz.  The errors in the fitting parameters were estimated
by fitting simulated power spectra with Gaussian noise.

\begin{figure}
\epsfxsize=3.375 in \epsfbox{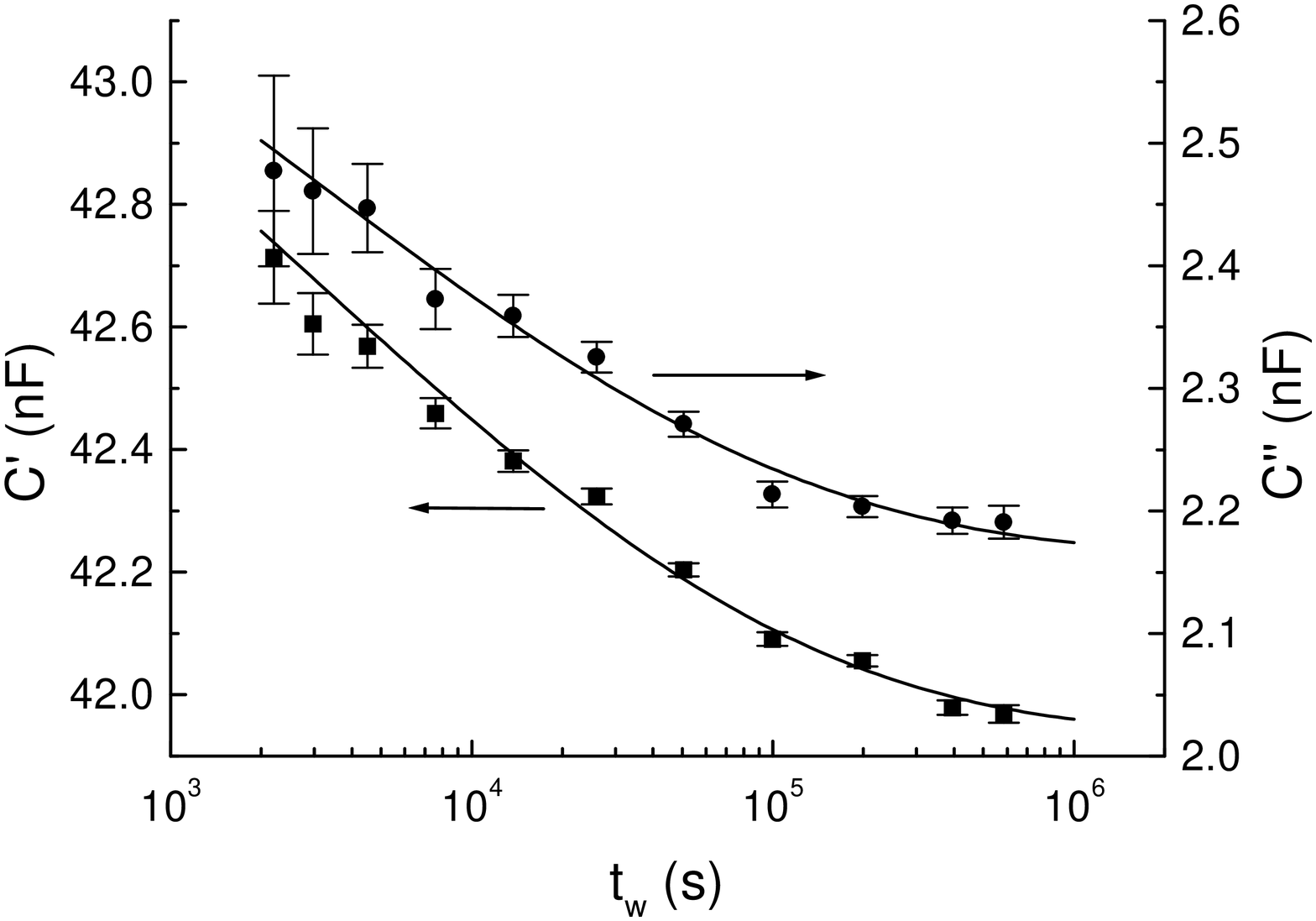}
\caption{Real (squares) and imaginary (circles)  capacitance
vs.\ waiting time, showing aging 
in the dielectric susceptibility at $T=179.8\,$K. Solid lines: fits
to stretched exponentials.}
\label{Caging}
\end{figure}

Figure~\ref{Caging} plots $C'$ and $C''$ vs.\ waiting time at
$T=179.8\,$K, which clearly show the aging of the dielectric
susceptibility, as previously seen by Leheny and Nagel\cite{leheny98}.
The curves can be fitted by a stretched exponential, $ C'(t_w) = C'_\infty
+ \Delta C' \exp \{- (t_w/\tau)^\beta \},$ with $\tau$ and $\beta$
taken from ref.\cite{leheny98}.

\begin{figure}
\epsfxsize=3.375 in \epsfbox{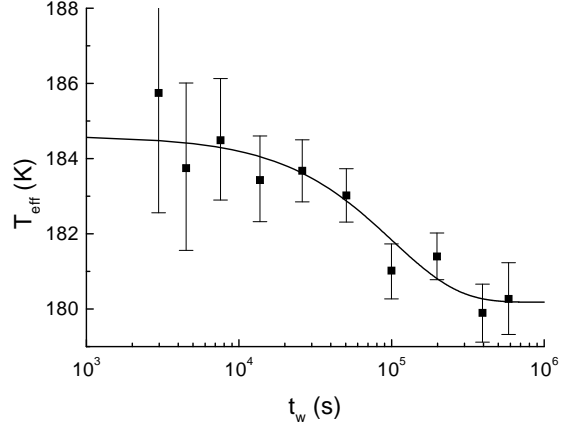}
\caption{Effective temperature $\Teff$ vs.\ waiting time for
$T=179.8\,$K. 
Points: experimental data, solid line: fit to exponential decay.}
\label{figTeff}
\end{figure}

\begin{figure}
\epsfxsize=3.375 in \epsfbox{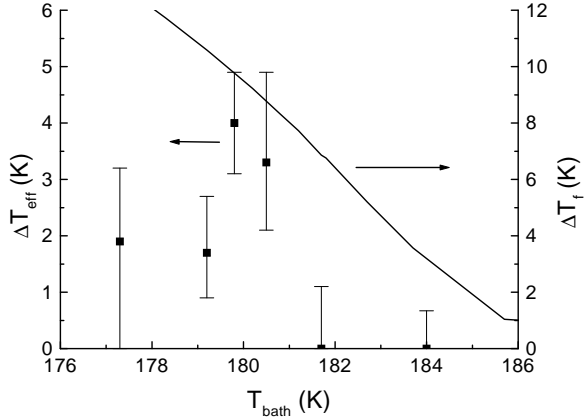}
\caption{$\Delta \Teff = \Teff(t_w=10000\,$s$) - \Teff(t_w=\infty)$ vs. 
bath temperature (points). Solid line is $T_f - T$.}
\label{DTeff}
\end{figure}

The effective temperature is plotted in fig.~\ref{figTeff} against
$t_w$ for $T=179.8\,$K. Two features of this graph are
noteworthy. First, a fit of the equilibrium data (long $t_w$) with
eq.~\ref{PSfit} yields the bath temperature with high
accuracy. Second, a fit of eq.~\ref{PSfit} to the non-equilibrium data
at early times {\em fails} to yield the bath temperature. This failure
to fit Nyquist's formula can {\em only} be explained by an FDT
violation. Although the violations are small, the statistical
certainty of a violation is very high: we performed the fitting
procedure on simulated spectra with the same susceptibility as the
experimental spectra but obeying the FDT. With $T=179.8\,$K, the
$\Teff$ obtained by fitting 500~000 simulated spectra was in {\em no
case} greater than $183.67\,$K, while 5 experimental points at the
earliest $t_w$ are above this value.  $\Teff$ relaxes slowly towards T
on a time scale similar to that of the aging in susceptibility.  The
decay can be fitted equally well by an exponential or a stretched
exponential. Most striking, this FDT violation persists much longer
than expected, to $\omega t_w > 10^5$.  This effect was clearly
observed only in a narrow range of temperatures around $180\,$K
(fig.~\ref{DTeff}), while susceptibility aging was seen over the range
$177\,$K--$184\,$K, in agreement with \cite{leheny98}.

It is interesting to compare and contrast the new results for $\Teff$
with other thermodynamic measurements. For example, dynamic heat
capacity measurements in glycerol\cite{birge} show that changes in
temperature are accompanied by heat flow on long time scales similar
to those found in other dynamical measurements.  There are two ways
this could be relevant to the present experiments.

First, heat flow will lead to a temperature gradient across the
sample, and thereby an apparent rise in $\Teff$. However, if this was
important, the susceptibility at early times should correspond to that
of higher temperatures in equilibrium, but we find that
$C'(179.8\,$K$)$ is at all $t_w$ lower than the final value of
$C'(181.7\,$K$)$.  Furthermore, a generous estimate of the
maximum temperature rise gives $50\,$mK at $180\,$K, negligible
compared with the measured rise in $\Teff$.

Second, the dynamic heat capacity measurements can be
understood\cite{simon97} in terms of an excess enthalpy, $\Delta H$,
trapped in the sample for times comparable to the relaxation time.  A
fictive temperature, $T_f$, \cite{tool} can be defined as the
temperature at which the glass and liquid lines intersect in a diagram
of enthalpy vs.\ temperature. Then $T_f = T + \Delta H/ \Delta C$,
where $\Delta C = C_g - C_l$ is the heat capacity difference between
the liquid and the glass\cite{simon97}. As the glass relaxes, its
enthalpy tends to that of the equilibrium liquid, and so $T_f \to
T$. It is tempting to attribute the excess $\Teff$ observed simply to
the excess enthalpy per CDF.  We can use $T_f - T$ as a rough estimate
of the excess enthalpy per CDF stored in the glass as a function of
$t_w$. We computed $T_f$ as described in \cite{simon97} using the
particular thermal history of our sample.  As fig.~\ref{DTeff} shows,
the deviation of $\Teff$ and $T_f$ from $T$ is similar in magnitude at
$180\,$K, but the temperature dependences of the two are distinctly
different.

That $\Teff$ and $T_f$ behave differently is not unexpected, since the
oscillator only couples to those CDF with relaxation time $\tau \sim
1/\omega \sim 0.1\,$s.  In a parallel relaxation picture, excess
enthalpy trapped in these relatively fast CDF would relax in $\sim
0.1\,$s and independently of slower CDF, too fast for this experiment
to detect.  It is possible that step-like relaxation of slow CDF near
the $\alpha$ peak could contribute enough to be observed at the much
higher measurement frequencies, as in Barkhausen
noise\cite{barkhausen}. Barkhausen noise decreases with frequency as
$1/\omega^2$, whereas the equilibrium noise decreases more slowly.
This would require a highly improbable $\Teff$ of $10^5\,$K at the
$\alpha$ peak frequency to increase $\Teff$ by $4\,$K at the
measurement frequency. Moreover, Monte Carlo simulations\cite{oursim}
of a parallel kinetics model with the same relaxation times as our
sample show no significant increase in the noise power at the
measurement frequency during a quench. Thus our results are
inconsistent with parallel kinetics.

Previous results, such as the aging of dielectric
response\cite{leheny98} and dielectric hole burning\cite{schiener}
could be qualitatively understood in terms of a slow relaxation of
relaxation rates via coupling between barriers and local density.  But
the present results further require that the CDF under observation be
continually `heated' above the phonon temperature. Hence some
mechanism must transfer excess energy stored in slower CDF to faster
CDF active in the experimental window. This could happen through
reshuffling of relaxation times of individual CDF, or by directed
energy exchange between different CDF.  The peak in figure~\ref{DTeff}
then arises naturally, because at high $T$ the relaxation is too fast
to be seen, while at low $T$ the energy release from relaxing CDF is
too slow.  Some form of series kinetics\cite{skin} (as seen in recent
simulations\cite{donati}), in which relaxation occurs in a series of
constrained steps, may explain these results.

In conclusion, we have reported experimental observation of weak FDT
violations in glycerol, which we have analyzed in terms of an
effective temperature. Like other quantities, the effective
temperature ages, but provides different information than fictive
temperature, susceptibility, or dynamic heat capacity. Specifically,
the present results indicate that successful models of the glass
transition must explicitly account for nontrivial energy exchange
between slower and faster configurational degrees of
freedom. Extending these results to a broader range of temperature,
frequency, and other materials would be useful.

We thank L.~F.~Cugliandolo and M.~B.~Weissman for helpful
discussions. This work was supported by NSF DMR 94-58008 and 98-77069,
and ACS-PRF 33329-AC5.  TSG was partly supported by Fundaci\'on
Antorchas (Argentina).

\end{document}